\documentclass[11pt]{article}
\usepackage{fullpage}
\usepackage{graphicx}
\usepackage{latexsym,amsmath,amssymb,amsthm,epic,eepic,multirow}
\usepackage{natbib}

\usepackage{multicol}
\usepackage{multirow}
\usepackage{subcaption} % for two figures side-by-side
\usepackage{amsfonts}
\usepackage{natbib}
\usepackage{algorithm}
\usepackage[noend]{algpseudocode}
\usepackage{mdframed}
\usepackage{tikz}
\usetikzlibrary{positioning}
\usepackage{hyperref}
\usepackage{listings} % for highlighting the R code
\usepackage{color} % for highlighting the R code
\usepackage{bbm}
\usepackage{relsize}
\usepackage{array}
\usepackage{mathtools}
\newcolumntype{P}[1]{>{\centering\arraybackslash}p{#1}}
\newcolumntype{R}[1]{>{\raggedleft\arraybackslash}p{#1}}

\definecolor{upmaroon}{rgb}{0.48, 0.07, 0.07}
\definecolor{royalazure}{rgb}{0.0, 0.22, 0.66}
\definecolor{pakistangreen}{rgb}{0.0, 0.4, 0.0}

\newcommand{\PP}{\text{pr}}
\newcommand{\EE}{E}
\newcommand{\eps}{\epsilon}

\theoremstyle{definition}
\newtheorem{theorem}{Theorem}%[section]
\newtheorem{proposition}{Proposition}
\newtheorem{lemma}{Lemma}%[section]
\newtheorem{corollary}{Corollary}

\newtheoremstyle{dotless}{}{}{}{}{\bfseries}{}{ }{}
\theoremstyle{dotless}

\usepackage{color}

\let\originalleft\left
\let\originalright\right
\renewcommand{\left}{\mathopen{}\mathclose\bgroup\originalleft}

\renewcommand{\right}{\aftergroup\egroup\originalright}
\makeatletter
\newcommand{\leqnomode}{\tagsleft@true}
\newcommand{\reqnomode}{\tagsleft@false}
 
\makeatother
\begin{document}
\title{Ancestor regression in linear structural equation models} 

\author{Christoph Schultheiss and Peter B\"uhlmann\\% and Ming Yuan\\
Seminar for Statistics, ETH Z\"urich}% and Department of Statistics, Columbia University}
%\date{}

\maketitle

\begin{abstract}
We present a new method for causal discovery in linear structural equation models. We propose a simple ``trick'' based on statistical testing in linear models that can distinguish between ancestors and non-ancestors of any given variable. Naturally, this can then be extended to estimating the causal order among all variables. %Unlike many methods, 
We provide explicit error control for false causal discovery, at least asymptotically. This holds true even under Gaussianity, where other
methods fail due to non-identifiable structures. These type I error guarantees come at the cost of reduced empirical power. Additionally, we provide an asymptotically valid goodness of fit p-value to assess whether multivariate data stems from a linear structural equation model.
\end{abstract}

\noindent%

\section{Introduction} \label{intro}
We propose a very simple yet effective method to infer the ancestor variables in a linear structural equation model from observational data. 
\begin{description}
\item \hspace*{6mm}
Consider a response variable of interest $Y$ and covariates $X$ in a linear structural equation model. The procedure is as follows. For a nonlinear function $f\left(\cdot\right)$, for example $f\left(Y\right) = Y^3$, run a least squares regression of $f(Y)$ versus $Y$ and all covariates $X$: 
 the p-value corresponding to the $k$-th covariate $X_k$ is measuring the significance that $X_k$ is an ancestor variable of $Y$, and it provides type I error control. 
\end{description}
We refer to this methods as ancestor regression. 
Its power (i.e., type II error) depends on the nature of the underlying data-generating probability distribution. 
Obviously, the proposed method is extremely simple and easy to be used; yet, it deals with the difficult problem of finding the causal order among random variables. In particular, the proposed method does not need any new software and it is computationally very efficient.

Structure search methods based on observational data for the graphical structure in linear structural equation models have been developed extensively for various settings: for the Markov equivalence class in linear Gaussian structural equation models \citetext{\citealp[Chapter~5.4]{spirtes2001causation}; \citealp{chickering2002optimal}} or for the single identifiable directed acyclic graph in non-Gaussian linear structural equation models \citep{shimizu2006linear, gnecco2021causal} or for models with equal error variances \citep{peters2014identifiability}. None of the methods comes with p-values and type I error control. In addition, for the identifiable cases, the corresponding algorithms require certain assumptions such as non-Gaussian errors. Particularly, the method from \cite{shimizu2006linear} and extensions thereof are not consistent when there are at least two normally distributed additive error terms involved such that false causal claims cannot be avoided even in the large sample limit. If the errors are just slightly non-Gaussian, the method requires very many samples to achieve a favorable behavior. In contrast, our procedure does not rely on any condition apart from linearity, but automatically exploits whether the structure is identifiable or not. In the latter case, we miss out on some causal relationships but our type I error control retains the same asymptotic guarantees. The price to pay for these guarantees is a reduced empirical power compared to competing methods, sometimes being substantial.

Regarding notation, we use upper case letters to denote a random variable, e.g., $X$ or $Y$. 
We use lower case letters to denote i.i.d.\ copies of a random variable, e.g., $x$. If $X \in \mathbb{R}^{p}$, then $x \in \mathbb{R}^{n \times p}$.
With a slight abuse of notation, $x$ can either denote the copies or realizations thereof.
We write $x_j$ to denote the j-th column of matrix $x$ and $x_{i,j}$ to denote the element in row $i$ and column $j$. With $\leftarrow$, we emphasize that an equality between random variables is induced by a causal mechanism. All proofs are given in Section \ref{app:proof} in the supplementary material.

\section{Ancestor regression}
\subsection{Model and method}\label{mandm}
Let $X \in \mathbb{R}^p$ be given by the following linear structural equation model
\begin{equation}\label{eq:SEM-model}
X_j \leftarrow \Psi_j + \sum_{k \in \text{PA}\left(j\right)} \theta_{j,k} X_k \quad j=1,\ldots ,p,
\end{equation}
where the $\Psi_1,\ldots ,\Psi_p$ are independent and centered random variables. We assume that $0 < \text{var}\left(\Psi_j\right) = \sigma_j^2 < \infty \ \forall j$ such that the covariance matrix of $X$ exists and has full rank. We use the notation $\text{PA}\left(j\right)$, $\text{CH}\left(j\right)$, $\text{AN}\left(j\right)$ and $\text{DE}\left(j\right)$ for $j$'s parents, children, ancestors, and descendants. Further, assume that there exists a directed acyclic graph (DAG) representing this structure.

Let $X_j$ with $j \in \left\{1, \ldots,p \right\}$ be a variable of interest; it has been denoted as response $Y$ in Section \ref{intro}. Consider a nonlinear function $f\left(\cdot\right)$. The following result describes the population property of ancestor regression, with general function $f\left(\cdot\right)$.

\begin{theorem}\label{theo:beta-fols}
Assume that the data $X$ follows the model \eqref{eq:SEM-model}. Consider the ordinary least squares regression $f\left(X_j\right)$ versus $X$, denote the according OLS parameter by $\beta^{f,j} \coloneqq \EE\left(XX^\top\right)^{-1}\EE\left\{Xf\left(X_j\right)\right\}$ and assume that it exists. Then,
\begin{equation*}
\beta_k^{f,j} = 0 \ \forall k \not \in \left\{\text{AN}\left(j\right)\cup j\right\}.
\end{equation*}
\end{theorem}
Importantly, $X_j$ itself must also be included in the set of predictors. The beauty of Theorem \ref{theo:beta-fols} lies in the fact that no assumptions on the distribution of the $\Psi_l$ or the size of the $\theta_{l,k}$, apart from existence of the moments, must be taken for any $l$ and $k \in \left\{1,\ldots ,p\right\}$. This allows one to control against false discovery of ancestor variables. 
%The proof of this and the subsequent theoretical results is available in Section \ref{app:proof} of the supplementary material.

Typically, $\beta_k^{f,j} \neq 0 $ holds for ancestors since a nonlinear function of that ancestor cannot be completely regressed out by the other regressors using only linear terms. For ancestors that are much further upstream, this effect might become vanishingly small. However, this is not such an issue since when fitting a linear model using the detected ancestors, those indirect ancestors are assigned a direct causal effect of $0$ anyway.

Based on Theorem \ref{theo:beta-fols}, we suggest testing for $\beta_k^{f,j} \neq 0$ in order to detect some or even all ancestors of $X_j$. Doing so for all $k$, requires nothing more than fitting a multiple linear model and using its corresponding
z-tests for individual covariates.

Let $x \in \mathbb{R}^{n \times p}$ be $n$ i.i.d.\ copies from the model \eqref{eq:SEM-model}. Define the following quantities
\begin{equation}\label{eq:hat-def}
\begin{aligned}
\hat{\beta}^{f,j} \coloneqq & \left(x^\top x\right)^{-1}x^\top f\left(x_j\right), \quad \hat{\sigma}^2 \coloneqq \dfrac{\left\Vert f\left(x_j\right) - x\hat{\beta}^{f,j}\right\Vert_ 2^2}{n-p} \quad \text{and} \quad \widehat{\text{var}}\left(\hat{\beta}^{f,j}_k\right)=\left(x^\top x\right)_{k,k}^{-1}\hat{\sigma}^2,
\end{aligned}
\end{equation}
where $f\left(\cdot\right)$ is meant to be applied elementwise in $f\left(x_j\right)$.

\begin{theorem}\label{theo:norm}
Assume that the data $X$ follows the model \eqref{eq:SEM-model}, $\EE\left\{f\left(X_j\right)^2\right\}< \infty$, $\EE\left(X_k^4\right)< \infty \ \forall k$ and $\beta^{f,j}$ exists. Let $x$ be $n$ i.i.d\ copies thereof. Using the definitions from \eqref{eq:hat-def}, it then holds
\begin{align*}
\hat{\beta}^{f,j}_k  & = \beta^{f,j}_k + {\scriptstyle \mathcal{O}}_p\left(1\right), \quad \widehat{\text{var}}\left(\hat{\beta}^{f,j}_k\right) = \mathcal{O}_p\left(\dfrac{1}{n}\right) \quad \text{and}\\
z_k^j & \coloneqq \dfrac{\hat{\beta}^{f,j}_k }{\surd{\widehat{\text{var}}\left(\hat{\beta}^{f,j}_k\right)}}\overset{\mathbb{D}}{\to}\mathcal{N}\left(0,1\right) \ \forall k \not \in \left\{\text{AN}\left(j\right)\cup j\right\}.
\end{align*}
\end{theorem}
Due to this limiting distribution, we suggest testing the null hypothesis $H_{0,k \rightarrow j}: \ k \not \in \text{AN}\left(j\right)$ with the p-value
\begin{equation}\label{eq:z-test}
p_k^j = 2 \left\{1 - \Phi \left(\vert z_k^j\vert\right)\right\},
\end{equation}
where $\Phi \left(\cdot\right)$ denotes the cumulative distribution function of the standard normal distribution.

For ancestors, for which $\beta^{f,j}_k \neq 0$, the absolute z-statistic increases as $\surd{n}$. In typical setups, one can thus detect all ancestors.
Having found all ancestors, one could infer the parents with an ordinary least squares regression of $X_j$ versus $X_{\text{AN}\left(j\right)}$, using the $t$-test for assigning the significance of being a parental variable. Such a procedure might have poor error control for low sample sizes as it requires full power in the first step to detect all ancestors; we provide error control only for the estimated ancestral set.

The choice of $f\left(\cdot\right)$ has an impact on the constant in the growth of $z_k^j$ for ancestors. If the $\Psi_l$ are symmetric, any even function yields $\beta_k^{f,j} = 0 \ \forall k$. Therefore, odd functions should be used. In our simulations and the real data analysis, we use $f\left(X_j\right) = X_j^3$ as it is the simplest odd function that only invokes slightly higher moments than linear functions. This choice leads to empirically competitive performance relative to other candidates in our simulations.

\subsection{Adversarial setups} \label{adv-setup}
There are cases where $\beta^{f,j}_k \neq 0$ does not hold true for some ancestors leading to reduced power of the method. We provide necessary and sufficient conditions for this and present examples. Define first the $j$-restricted Markov boundary of $k$ to be
\begin{equation*}
\text{MA}^{\rightarrow j}\left(k\right) \coloneqq \left[\text{PA}\left(k\right) \cup \text{CH}\left(k\right) \cup \underset{l \in \text{CH} \left(k\right)}{\bigcup}\left\{\text{PA}\left(l\right)\setminus k\right\}\right] \cap \left\{\text{AN}\left(j\right)\cup j \right\}.
\end{equation*}
It contains all the variables in the Markov boundary of $k$ which are ancestors of $j$ or $j$ itself. E.g., if $k \in \text{AN}\left(j\right)$ all its parents are in the restricted Markov boundary, but not necessarily all its children.
\begin{theorem}\label{theo:adv}
Let $k \in \text{AN}\left(j\right)$. 
Then, 
\begin{equation*}
\beta_k^{f,j}=0 \quad \forall f\left(\cdot\right) \quad \text{if and only if} \quad E\left(X_k \mid X_j\right) =  E \left(X_{\text{MA}^{\rightarrow j}\left(k\right)}^\top \gamma^{j,k} \mid X_j\right),
\end{equation*}
where $\gamma^{j,k}$ is the least squares parameter for regressing $X_k$ versus $X_{\text{MA}^{\rightarrow j}\left(k\right)}$. In particular,
\begin{equation*}
\beta_k^{f,j}=0 \quad \forall f\left(\cdot\right) \quad \text{if} \quad E\left(X_k \mid X_{\text{MA}^{\rightarrow j}\left(k\right)}\right) =  X_{\text{MA}^{\rightarrow j}\left(k\right)}^\top \gamma^{j,k}.
\end{equation*}
\end{theorem}
Intuitively speaking, if the conditional expectation of $X_k$ given the $j$-restricted Markov boundary is linear, $k$ could also be a child of all these variables. Thus, it is not detectable as ancestor of $j$. In the following, we present two examples that fulfil the conditions of Theorem \ref{theo:adv}. These are the only examples we know of.

\paragraph{Gaussian $\Psi$.}
It is well-known in causal discovery for linear structural equation models that Gaussian error terms lead to non-identifiability.

Define
\begin{equation*}
\text{CH}^{\rightarrow j}\left(k\right) \coloneqq \left[\text{CH}\left(k\right) \cap \left\{ \text{AN}\left(j\right)\cup j\right\}\right],
\end{equation*}
i.e., the children of $k$ through which a directed path from $k$ to $j$ begins.
\begin{proposition}\label{prop:gauss-psi}
Assume that the data $X$ follows the model \eqref{eq:SEM-model}. Let $k \in \text{AN}\left(j\right)$ with $\Psi_k \sim \mathcal{N}\left(0,\sigma_k^2\right)$.  Then, it holds
\begin{equation*}
\beta_k^{f,j} = 0 \quad \forall f\left(\cdot\right) \quad \text{if} \quad \Psi_l \sim \mathcal{N}\left(0,\sigma_l^2\right) \ \forall l \in \text{CH}^{\rightarrow j}\left(k\right) .
\end{equation*}
Under the additional assumptions of Theorem \ref{theo:norm},
\begin{equation*}
z_k^j \coloneqq \dfrac{\hat{\beta}^{f,j}_k }{\surd{\widehat{\text{var}}\left(\hat{\beta}^{f,j}_k\right)}}\overset{\mathbb{D}}{\to}\mathcal{N}\left(0,1\right).
\end{equation*}
\end{proposition}
Thus, if every directed path from $k$ to $j$ starts with an edge for which the nodes on both ends have Gaussian noise terms, we have no power to detect this ancestor relationship. However, we neither detect the opposite direction as guaranteed by Theorem \ref{theo:beta-fols}, and thus, control against false positives is guaranteed.

\paragraph{Special constellation of distributions and coefficients.}    
A pathological case occurs if a child's, say, $l$, error term has the same distribution as the inherited contribution from the parent's, say, $k$, error term. Then, $k$ is not detectable as $l$'s ancestor. Likewise, it is not detected as ancestor of any of $l$'s descendants $j$ to which all directed paths from $k$ start with the edge $k \rightarrow l$.
\begin{proposition}\label{prop:distr}
Assume that the data $X$ follows the model \eqref{eq:SEM-model}. Let $k \in \text{AN}\left(j\right)$ and $\text{CH}^{\rightarrow j}\left(k\right)=\{l\}$. Then, it holds
\begin{equation*}
\beta_k^{f,j} = 0 \quad \forall f\left(\cdot\right) \quad \text{if} \quad \Psi_l \overset{\mathbb{D}}{=} \theta_{l,k} \Psi_k. 
\end{equation*}
\end{proposition}
For the variables discussed here, the limiting Gaussian distribution as stated in Theorem \ref{theo:norm} is not guaranteed even though $\beta_k^{f,j} = 0$; see also the proof in the supplemental material.

\subsection{Simulation example}\label{sim:anc}
We study ancestor regression in a small simulation example. We generate data from a linear structural equation model with $6$ variables. The causal order is fixed to be $X_1$ to $X_6$. Otherwise, the structure is randomized and changes per simulation run: $X_k$ is a parent of $X_l$ for $k<l$ with probability $0.4$ such that there is an average of $6$ parental relationships. The edge weights are sampled uniformly and the $\Psi_k$ are assigned by permuting a fixed set of $6$ error distributions. The full data generating process can be found in Section \ref{app:sim} of the supplementary material.

We aim to find the ancestors of $X_4$ which can be any subset of $\left\{X_1,X_2,X_3\right\}$. We create $1000$ different setups and test each on sample sizes varying from $10^2$ to $10^6$. As a nonlinear function, we use $f\left(X_j\right) = X_j^3$. By z-statistic, we mean $z_k^4$ as defined in Theorem \ref{theo:norm}. We calculate p-values according to \eqref{eq:z-test} and apply a Bonferroni-Holm correction (without cutting off at $1$ for the sake of visualization) to them.

\begin{figure}[b!]
%\spacingset{1}
 \centering
 \includegraphics[width=0.8\textwidth]{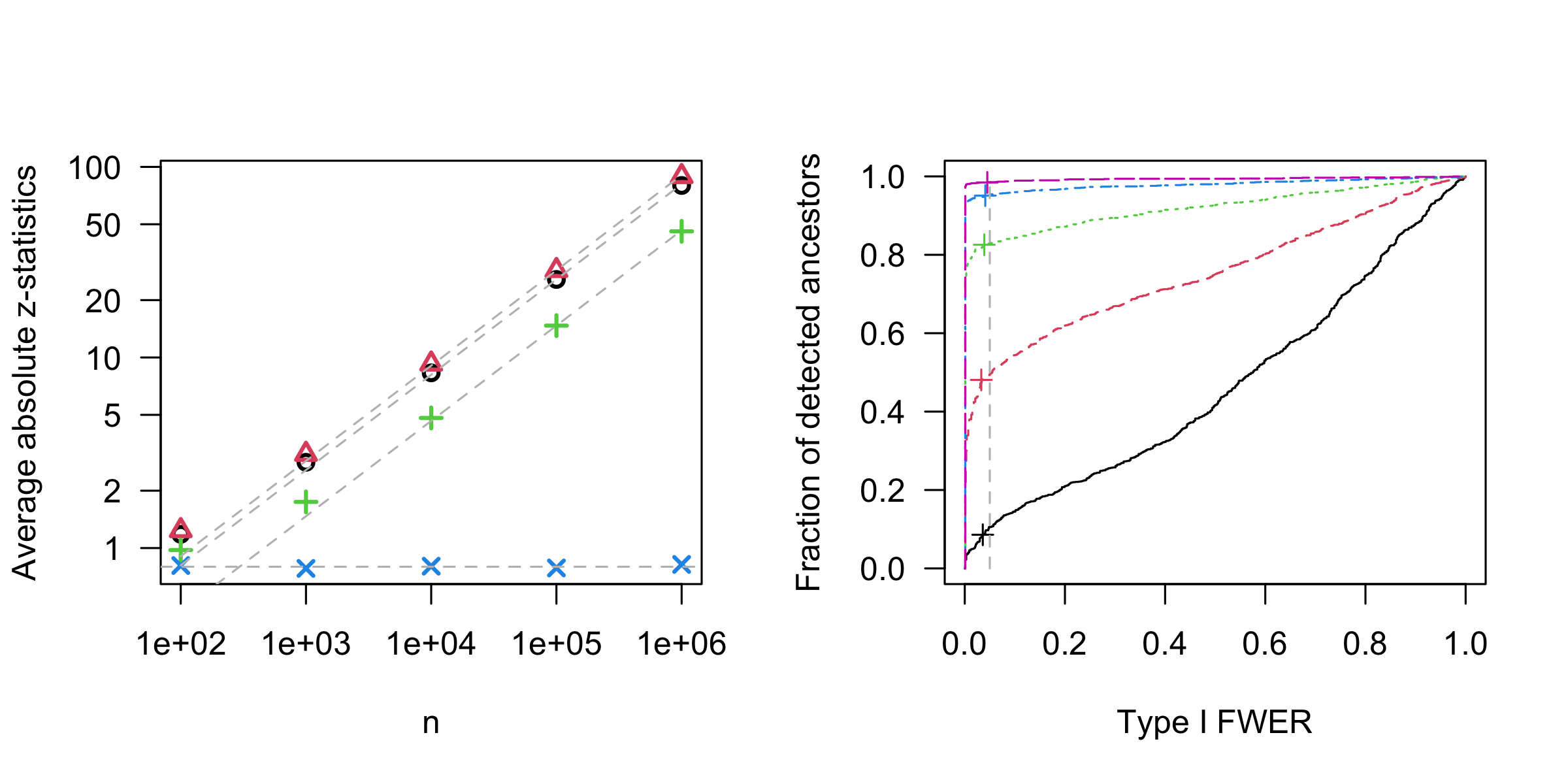}
 \caption[Simulation in a linear SEM]
 {Detecting the ancestors of $X_4$ in a linear structural equation model with $6$ variables. The results are based on $1000$ simulation runs. On the left: Average absolute z-statistic for all ancestors (circles, black), parents (triangles, red), non-parental ancestors (pluses, green), and non-ancestors (crosses, blue) for different sample sizes. The dashed diagonals correspond to $\surd{n}$-growth fitted to perfectly match at $n=10^5$. The horizontal line corresponds to $\left(2/\pi\right)^{1/2}$, i.e., the first absolute moment of the asymptotic null distribution, a standard Gaussian. On the right: fraction of simulation runs with at least one false causal detection versus fraction of detected ancestors for the different sample sizes $10^2$ (solid, black), $10^3$ (dashed, red), $10^4$ (dotted, green), $10^5$ (dot-dashed, blue), and $10^6$ (long-dashed, pink). The curve uses the level $\alpha$ of the test as implicit curve parameter. The pluses correspond to nominal $\alpha = 5\%$. The vertical line is at actual $5\%$.}
 \label{fig:anc-x4-rand}
\end{figure}

In Fig.\ \ref{fig:anc-x4-rand}, we see the desired $\surd{n}$-growth of the absolute z-statistic for the ancestors, while for the non-ancestors their sample averages are close to the theoretical mean under the asymptotic null distribution. Indirect ancestors are harder to detect than parents.
Although the null distribution is only asymptotically achieved, the type I family-wise error rate is controlled for every sample size, supporting our method's main benefit, i.e., robustness against false causal discovery. 

\section{Ancestor detection in networks: nodewise and recursive}
\subsection{Algorithm and goodness of fit test}\label{anc-alg}
In the previous section, we assumed that there is a (response) variable $X_j$ that is of special interest. This is not always the case. Instead, one might be interested in inferring the full set of causal connections between the variables. Naturally, our ancestor detection technique can be extended to that problem by applying it nodewise. We suggest the procedure sketched below. The detailed algorithm can be found in Section \ref{app:algo} of the supplementary material. Notably, the algorithm is invariant to the ordering of the variables.

First, the set of ancestors is defined based on the significant p-values, after multiplicity correction over all $p\left(p-1\right)$ z-tests, of ancestor regression. Any correction controlling the type I family-wise error rate is applicable, and we use here Bonferroni-Holm. Next, further ancestral relationships are constructed recursively by adding the estimated ancestors of every estimated ancestor. This recursive construction facilitates the detection of all ancestors. This procedure cannot increase the type I family-wise error rate compared to just using the significant p-values 
because a false causal discovery can only be propagated if it existed in the first place. 

Since there is no guarantee that the recursive construction does not create directed cycles, i.e., variables are claimed to be their own ancestors, we need to address this. 
If such cycles are found, the significance level is gradually reduced until no more directed cycles are outputted. This means that the output becomes somewhat independent of the significance level, e.g., in a case with two variables and $p_1^2=10^{-6}$ and $p_2^1=10^{-3}$ as in \eqref{eq:z-test}, we would never claim $X_2 \rightarrow X_1$ no matter how large $\alpha$ is chosen. We denote the estimated set of ancestors for $X_j$ by $\widehat{\text{AN}}\left(j\right)$. Notably, the algorithm determines a causal order between the variables but does not always lead to a unique parental graph. For instance, if $\widehat{\text{AN}}\left(3\right) = \left\{1, 2\right\}$ and $\widehat{\text{AN}}\left(2\right) = \left\{1\right\}$, $X_1$ might be a causal parent of $X_3$ but its effect could also be fully mediated by $X_2$.

One can consider the largest significance level such that no loops are created as a p-value for the null hypothesis that the modeling assumption \eqref{eq:SEM-model} holds true. We denote this level, which is a further output of our algorithm, by $\hat{\alpha}$. Thus, we provide a goodness of fit test for our modelling assumption with an asymptotically valid p-value: a small realized $\hat{\alpha}$ would provide evidence against the linear structural equation model in \eqref{eq:SEM-model}. If such evidence exists, it is advisable to take the outcome of ancestor regression or other causal discovery methods relying on linear structural equation models with a grain of salt. We make use of this p-value in the data analysis in Section \ref{sachs}. We summarize the properties of our algorithm.

\begin{corollary}\label{corr:algo}
Assume that the conditions of Theorem \ref{theo:norm} hold $\forall j \in \left\{1,\ldots,p\right\}$. Let $\widehat{\text{AN}}\left(j\right) \ \forall j \in \left\{1,\ldots,p\right\}$ and $\hat{\alpha}$ be the output of the nodewise ancestor regression algorithm with significance level $\alpha$ and Bonferroni-Holm correction. Then, it holds
\begin{equation*}
\underset{n \rightarrow \infty}{\text{lim}}\PP\left\{\exists j,k\neq j: \ k \not\in \text{AN}\left(j\right) \text{ and } k \in \widehat{\text{AN}}\left(j\right) \right\} \leq \alpha \quad \text{and} \quad \underset{n \rightarrow \infty}{\text{lim}}\PP\left(\hat{\alpha} \leq \alpha' \right) \leq \alpha' \ \forall \alpha' \in \left(0,\alpha\right).
\end{equation*}
\end{corollary}

\subsection{Simulation example}
We extend the simulation from Section \ref{sim:anc} to estimating the ancestors of each variable using the algorithm described in Section \ref{anc-alg}. We compare our method to LiNGAM \citep{shimizu2006linear} using the implementation provided in the \texttt{R}-package \texttt{pcalg} \citep{kalisch2012causal}. For every simulation run, we use two slighlty different data generating processes. In the first, only one of the $\Psi_k$ follows a Gaussian distribution, in the second, there are two error terms with normal distribution and an edge between the two respective nodes is always present. As LiNGAM provides an estimated set of parents, we additionally apply our recursive algorithm to the output to get an estimated set of ancestors which enables comparison with our method.

\begin{figure}[b!]
%\spacingset{1}
 \centering
 \includegraphics[width=0.8\textwidth]{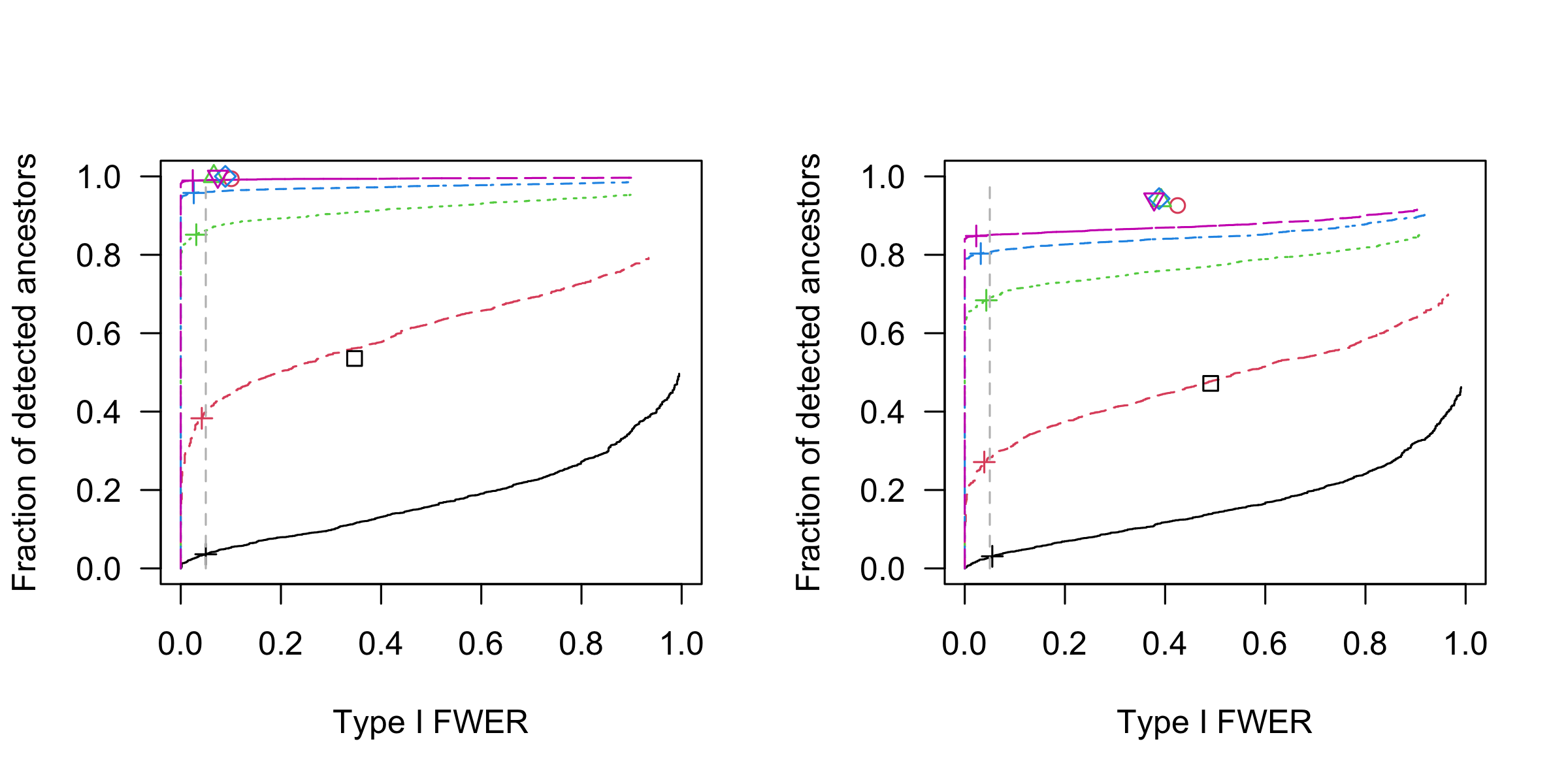}
 \caption[Simulation in a linear SEM]
 {Nodewise ancestor detection in a linear structural equation model with $6$  variables. The results are based on $1000$ simulation runs. Depicted is the family-wise error rate of false causal detection versus the fraction of detected ancestors. The curves use the level of the test $\alpha$ as implicit curve parameter. The pluses correspond to nominal $\alpha = 5\%$. The vertical line is at actual $5\%$. The other symbols correspond to the performance of the LiNGAM algorithm. We consider the different sample sizes $10^2$ (solid / square, black), $10^3$ (dashed / circle, red), $10^4$ (dotted / triangle pointing upward, green), $10^5$ (dot-dashed / diamond, blue), and $10^6$ (long-dashed / triangle pointing downward, pink). On the left: exactly $1$ error term follows a Gaussian distribution. On the right: exactly $2$ error terms follow a Gaussian distribution.}
 \label{fig:graph-perf-rand}
\end{figure}

The results are shown in Fig. \ref{fig:graph-perf-rand}. For the model with only one Gaussian error variable, we can reliably detect almost all ancestors without any false causal claims for large enough sample sizes. The few exceptions can be explained as some setups can be very close to the non-identifiable case discussed in Proposition \ref{prop:distr}. Not all curves reach a power of $1$ even when letting the significance level become arbitrarily large. This can be
explained by the possible insensitivity to the significance level, as sketched in Section \ref{anc-alg}.

We are able to control the family-wise error rate even for low sample size using a nominal size of $\alpha = 5\%$ supporting our theoretical results. This is not the case for LiNGAM. LiNGAM is designed such that it always must determine a causal order based on the underlying independent component analysis \citep{hyvarinen1999fast} even when sufficient information is not available. Therefore, no type I error guarantees can be provided. The power of LiNGAM approaches $1$ much faster than ancestor regression and if one allows for a bit more liberate type I error, LiNGAM  appears preferable in the model with one Gaussian noise term. The picture changes when looking at slight violations of the LiNGAM assumption, i.e., another Gaussian error term. LiNGAM is still more powerful but does not control the error at all. No matter the sample size, a wrong causal claim is made in around $40\%$ of the setups. Ancestor regression is more robust to this deviation as the type I error guarantees do not require non-Gaussian error terms. For the unidentifiable edges, it avoids making any decision and can control the error rate at any desired level at the price of some power reduction. In this simulation, Proposition \ref{prop:gauss-psi} applies to around $14\%$ of the ancestral connections.

We provide additional simulation results for settings varying between non-Gaussian and Gaussian scenarios in Section \ref{app:sim-mix} in the supplementary material. When being close to the fully Gaussian case, despite satisfying the LiNGAM assumption (Shimizu et al., 2006) in population, this clearly worsens the performance of LiNGAM for finite sample size. 

\section{Real data example}\label{sachs}
\begin{table}[b!]
%\spacingset{1}
\centering
%\subfloat[]{
\begin{tabular}{|P{0.2\textwidth}R{0.2\textwidth}R{0.2\textwidth}P{0.07\textwidth}P{0.07\textwidth}|}
\hline
Causal effect & ancestor regression & linear regression & SC & MH\\
\hline
PIP3 $\rightarrow$ PIP2 & 3.3e-39  &  5.5e-43 & $\longrightarrow$ & $\longrightarrow$\\
PIP3 $\rightarrow$ PLCg & 6.7e-39  &  1.4e-36 & $\longrightarrow$ & $\dashrightarrow$\\
PKA $\rightarrow$ Erk & 2.9e-26  &  7.2e-2 & $\longrightarrow$ & $\dashrightarrow$\\
JNK $\rightarrow$ p38 & 6.6e-20  &  2.4e-19 & --& --\\
PKA $\rightarrow$ Akt & 7.2e-20  &  9.4e-4 & $\longrightarrow$ & $\longrightarrow$\\
JNK $\rightarrow$ PKC & 1.2e-16  &  5.1e-88 & $\longleftarrow$ & $\longleftarrow$\\
RAF $\rightarrow$ MEK & 5.4e-15  &  0 & $\longrightarrow$ & $\longleftarrow$\\
PKC $\rightarrow$ p38 & 3.1e-13  &  0 & $\longrightarrow$ & $\longrightarrow$\\
Akt $\rightarrow$ Erk & 7.6e-07  &  0 & -- & $\longrightarrow$\\
\hline
\end{tabular}
\caption{Analysis of the dataset by \cite{sachs2005causal}.
The second column reports the raw p-value from ancestor regression, $p_k^j$, associated with this edge and the third column the raw p-value from the subsequent linear model fit. The rows are ordered by the p-value from ancestor regression from low to high. We present the conclusions of the consensus network in \cite{sachs2005causal} (column SC) and the method from \cite{mooij2013cyclic} (column MH): the edge is present ($\longrightarrow$), there exists a directed path with the same orientation but no edge ($\dashrightarrow$), the edge is reversed ($\longleftarrow$), there is no directed path (-).}
\label{tab:sachs-new}
\end{table}
We analyze the flow cytometry dataset presented by \citet{sachs2005causal}. It contains cytometry measurements of 11 phosphorylated proteins and
phospholipids. Data is available from various experimental conditions, some of which are interventional environments. The authors provide a ``ground truth'' on how these quantities affect each other, the so-called consensus network. The dataset has been further analyzed in various follow-up papers, see, e.g., \citet{mooij2013cyclic} and \citet{taeb2021perturbations}. Following these works, we consider data from 8 different environments, 7 of which are interventional. The sample size per environment ranges from $707$ to $913$.

For each environment individually, we estimate the ancestral relationships using our recursive algorithm sketched in Section \ref{anc-alg} with nonlinear function $f\left(X_j\right) = X_j^3$ and $\alpha = 0.05$. The goodness of fit p-value $\hat{\alpha}$ per environment, but corrected for the number of environments, ranges from $0.14$ to $3 \times 10^{-12}$. All but one p-value are lower than $0.04$, indicating for these environments that the data does not follow the model \eqref{eq:SEM-model}. The deviation can be in terms of hidden variables, nonlinear effects, or noise that is not additive. While the before mentioned and other published findings usually result in one graph harmonized over different environments, our highly varying results across environments suggest to question a standard ``autonomy assumption'' in causality \citep{aldrich1989autonomy}  that an intervention does not change the underlying graph (except
for edges that point into the intervened node).

Subsequently, we focus on the environment with the highest $\hat{\alpha}$ which seems to be most conformable with a linear structural equation model. The dataset contains $723$ observations. For each node, we fit a linear model using the claimed set of ancestors as predictors to see which ancestors might be direct parents.
We summarize our findings in Table \ref{tab:sachs-new}. Most ancestors show indication of being direct parents. However, as laid out in Section \ref{mandm}, we do not have type I error guarantees for finding parents in case some ancestors are missing. 

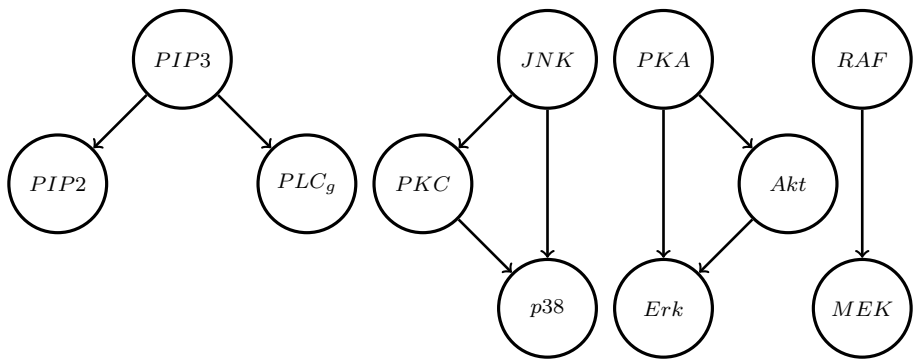
\begin{figure}[t!]
\centering
\begin{tikzpicture}[roundnode/.style={circle, very thick, minimum size=13mm}]
% nodes %
\node[draw, roundnode, text centered] (pip3) {${\scriptstyle PIP3}$};
\node[draw, roundnode, below left =1cm of pip3, text centered] (pip2) {${\scriptstyle PIP2}$};
\node[draw, roundnode, below right =1cm  of pip3, text centered] (plcg) {${\scriptstyle PLC_g}$};
\node[draw, roundnode, right =0.2 cm of plcg, text centered] (pkc) {${\scriptstyle PKC}$};
\node[draw, roundnode, above right = 1cm of pkc] (jnk) {${\scriptstyle JNK}$};
\node[draw, roundnode, below right = 1cm of pkc, text centered] (p38) {${\scriptstyle p38}$};
\node[draw, roundnode, right = 0.2cm of jnk] (pka) {${\scriptstyle PKA}$};
\node[draw, roundnode, below right =1cm of pka, text centered] (akt) {${\scriptstyle Akt}$};
\node[draw, roundnode, below left = 1cm of akt, text centered] (erk) {${\scriptstyle Erk}$};
\node[draw, roundnode, right = 1.3cm of pka] (raf) {${\scriptstyle RAF}$};
\node[draw, roundnode, right  = 1.3cm of erk, text centered] (mek) {${\scriptstyle MEK}$};

% edges %
\draw[->, line width= 1] (pip3) -- (pip2);
\draw[->, line width= 1] (pip3) -- (plcg);
\draw[->, line width= 1] (jnk) -- (pkc);
\draw[->, line width= 1] (jnk) -- (p38);
\draw[->, line width= 1] (pkc) -- (p38);
\draw[->, line width= 1] (pka) -- (akt);
\draw[->, line width= 1] (pka) -- (erk);
\draw[->, line width= 1] (akt) -- (erk);
\draw[->, line width= 1] (raf) -- (mek);
\end{tikzpicture} 
 \caption[DAG]
 {Ancestral relations for flow cytometry data obtained with ancestor regression}
 \label{fig:Sachs-DAG}
\end{figure}

For comparison, we show what conclusion the consensus network as well as \cite{mooij2013cyclic} draw for these edges. Our method is in agreement with at least one of these works except for the two edges coming from JNK. One of the indirect paths in \cite{mooij2013cyclic} corresponds to the highest p-value in the linear model fit, which is a further agreement. Our outputted ancestral graph, see Figure \ref{fig:Sachs-DAG}, consists of $4$ disconnected components.
When considering these components
individually, we note that the part containing JNK, where we receive somewhat unexpected findings, has the strongest indication of violating the model assumptions in terms of the goodness of fit p-value $\hat{\alpha}$.

\section*{Acknowledgment}
The project leading to this application has received funding from the European Research Council (ERC) under the European Union’s Horizon 2020 research and innovation programme (grant agreement No 786461).

\bibliographystyle{apalike} 
\bibliography{references}

\newpage
\appendix
\allowdisplaybreaks

\newpage
\section{Proofs}\label{app:proof}
\subsection{Additional notation}
We introduce additional notation that is used for these proofs.

Subindex $-k$, e.g., $x_{-k}$ denotes a matrix with all columns but the $k$-th.
$I_n$ is the $n$-dimensional identity matrix.
$P_{-k}$ denotes the orthogonal projection onto $x_{-k}$ and $P^{\perp}_{-k} = I_n - P_{-k}$ denotes the orthogonal projection onto its complement. $P_{x}$ is the orthogonal projection onto all $x$. 

For some random vector $X$, we have the moment matrix $\Sigma^{X} \coloneqq \EE\left(XX^\top\right)$. This equals the covariance matrix for centered $X$. We assume this matrix to be invertible. Then, the principal submatrix $\Sigma^X_{-j,-j} \coloneqq \EE\left(X_{-j}X_{-j}^\top\right)$ is also invertible. We denote statistical independence by $\perp$.
\subsection{Previous work}
We adapt some definitions from and results proven in \citet{schultheiss2022assessing}.
\begin{equation} \label{eq:z-w-def}
 \begin{aligned}
Z_k & \coloneqq X_k - X_{-k}^{\top}\gamma_k, \quad &&\text{where}\\
\gamma_k & \coloneqq \underset{b \in \mathbb{R}^{p-1}}{\text{argmin}}\EE\left\{\left(X_k-X_{-k}^\top b\right)^2\right\} = \left(\Sigma^X_{-k,-k}\right)^{-1}\EE\left(X_{-k}X_k\right),\\
W_k & \coloneqq f\left(X_j\right) - X_{-k}^{\top}\zeta_k, \quad &&\text{where}\\
 \zeta_k & \coloneqq \underset{b \in \mathbb{R}^{p-1}}{\text{argmin}}\EE\left\{\left(f\left(X_j\right)-X_{-k}^\top b\right)^2\right\} = \left(\Sigma^X_{-k,-k}\right)^{-1}\EE\left\{X_{-k}f\left(X_j\right)\right\}.
 \end{aligned}
\end{equation}
Using these definitions, we have $\beta_k^{f,j} = \EE\left(Z_k W_k\right)/\EE\left(Z_k^2\right) = \EE\left\{Z_k f\left(X_j\right)\right\}/\EE\left(Z_k^2\right)$ from partial regression. We cite a Lemma fundamental to our results.
\begin{lemma}\label{lemm:zrep}
Assume that the data follows the model \eqref{eq:SEM-model} without hidden variables. Then,
\begin{equation*}
Z_k = \delta_{k,k} \Psi_k + \sum_{l \in \text{CH}\left(k\right)} \delta_{k,l} \Psi_l \quad k=1,\ldots ,p
\end{equation*}
for an appropriate set of parameters. Further, the support of $\gamma_j$ (cf.\ \eqref{eq:z-w-def}) is restricted to $j$'s Markov boundary.
\end{lemma}
\subsection{Proof of Theorem \ref{theo:beta-fols}}
Let $\Psi = \left(\Psi_1, \ldots, \Psi_p\right)^\top$. Then, we can write $X = \omega\Psi$ for a suitable $\omega$ with $\omega_{lk} = 0$ if $l \not \in \left\{\text{DE}\left(k\right)\cup k \right\}$.
We can now find $\beta^{f,j}$ using this representation.
\begin{align*}
\beta^{f,j} & = \EE\left(XX^\top\right)^{-1} \EE\left\{Xf\left(X_j\right\}\right)=\left(\omega^{-1}\right)^\top \EE\left(\Psi\Psi^\top\right)^{-1}\omega^{-1}\omega\EE\left\{\Psi f\left(X_j\right)\right\} \\
&=\left(\omega^{-1}\right)^\top \text{diag}\left\{\dfrac{1}{\EE\left(\Psi_1^2\right)},\ldots,\dfrac{1}{\EE\left(\Psi_p^2\right)}\right\}\EE\left\{\Psi f\left(X_j\right)\right\}=\left(\omega^{-1}\right)^\top \left[\dfrac{\EE\left\{\Psi_1 f\left(X_j\right) \right\}}{\EE\left(\Psi_1^2\right)},\ldots,\dfrac{\EE\left\{\Psi_p f\left(X_j\right) \right\}}{\EE\left(\Psi_p^2\right)}\right]^\top .
\end{align*}
The third equality follows from the independence of the $\Psi_l$.
Naturally, for all $l \not \in \left\{\text{AN}\left(j\right)\cup j\right\}$ we have $\Psi_l \perp X_j$ such that $\EE\left\{\Psi_l f\left(X_j\right) \right\}=0$. Further, $\omega^{-1}_{lk}=0$ if $l \not \in \left\{\text{DE}\left(k\right)\cup k \right\}$. To see this, note that $\omega$ would be lower triangular, if $1,\ldots, p$ denoted a causal order. Then, its inverse would be lower triangular as well. Naturally, the same principle applies for every other permutation. Thus,
\begin{equation*}
\beta_k^{f,j} =\sum_{l}{\omega^{-1}_{lk}\dfrac{\EE\left\{\Psi_l f\left(X_j\right) \right\}}{\EE\left(\Psi_l^2\right)}}=\sum_{l \in \left\{\text{DE}\left(k\right)\cup k \right\}}{\omega^{-1}_{lk}\dfrac{\EE\left\{\Psi_l f\left(X_j\right) \right\}}{\EE\left(\Psi_l^2\right)}}=\sum_{l \in \left[ \left\{\text{DE}\left(k\right)\cup k \right\}\cap\left\{\text{AN}\left(j\right)\cup j \right\}\right]}{\omega^{-1}_{lk}\dfrac{\EE\left\{\Psi_l f\left(X_j\right) \right\}}{\EE\left(\Psi_l^2\right)}}
\end{equation*}
such that $\beta_k^{f,j}=0$ if $\left\{\text{DE}\left(k\right)\cup k \right\}\cap\left\{\text{AN}\left(j\right)\cup j \right\}= \emptyset$, i.e., if $k$ is not an ancestor of $j$.

Alternatively, we could invoke Lemma \ref{lemm:zrep} to see that $Z_k \perp X_j$ for a non-ancestor $k$. Then, $\beta_k^{f,j} = \EE\left\{Z_k f\left(X_j\right)\right\}/\EE\left(Z_k^2\right)=0$

\subsection{Proof of Theorem \ref{theo:norm}}
Define 
\begin{equation*}
f\left(X_j\right) \coloneqq X^\top \beta^{f,j} + \mathcal{E}, \quad \hat{z}_k = P^{\perp}_{-k} x_k \quad \text{and} \quad \hat{w}_k = P^{\perp}_{-k} f\left(x_j \right) \quad \text{such that} \quad \hat{\beta}_k^{f,j} = \dfrac{\hat{z}_j^\top \hat{w}_j}{\hat{z}_j^\top \hat{z}_j}.
\end{equation*}
Since we assume the covariance matrix to be bounded, we find
\begin{align*}%\label{eq:conv-inv}
\begin{split}
\dfrac{1}{n}x_{-k}^\top x_{-k}\overset{\mathbb{P}}{\to} \Sigma^X_{-k,-k} & \implies n \left(x_{-k}^\top x_{-k}\right)^{-1} \overset{\mathbb{P}}{\to} \left(\Sigma^X_{-k,-k}\right)^{-1} \\
 & \implies \left\Vert n \left(x_{-k}^\top x_{-k}\right)^{-1} \right\Vert \overset{\mathbb{P}}{\to} \left\Vert \left(\Sigma^X_{-k,-k}\right)^{-1} \right\Vert = \mathcal{O}\left(1\right),
\end{split}
\end{align*}
where we use invertibility and the continuous mapping theorem. This then implies
\begin{align*}%\label{eq:P_j-conf}
\begin{split}
\vert z_k ^\top P_{-k} w_k \vert & = \vert z_k^\top x_{-k} \left( x_{-k}^\top x_{-k}\right)^{-1}x_{-k}^\top w_k \vert \leq \left\Vert z_k^\top x_{-k} \right\Vert_2 \left\Vert \left( x_{-k}^\top x_{-k}\right)^{-1} \right\Vert_2 \left\Vert x_{-k}^\top w_k \right\Vert_2 \\ 
& \leq \left\Vert z_k^\top x_{-k} \right\Vert_1 \left\Vert \left( x_{-k}^\top x_{-k}\right)^{-1} \right\Vert_2 \left\Vert x_{-k}^\top w_k \right\Vert_1 = \sum_{l \neq k } \vert z_k^\top x_l \vert \left\Vert \left( x_{-k}^\top x_{-k}\right)^{-1} \right\Vert_2 \sum_{l \neq k } \vert x_l^\top w_k \vert \\
& = \mathcal{O}_p\left(\surd{n}\right)\mathcal{O}_p\left(\dfrac{1}{n}\right){ \scriptstyle \mathcal{O}}_p\left(n\right)= { \scriptstyle \mathcal{O}}_p\left(\surd{n}\right)
\end{split}
\end{align*}
and analogously
\begin{equation*}
\vert z_k ^\top P_{-k} z_k \vert = \mathcal{O}_p\left(\surd{n}\right)\mathcal{O}_p\left(\dfrac{1}{n}\right)\mathcal{O}_p\left(\surd{n}\right) = \mathcal{O}_p\left(1\right).
\end{equation*}
We get a better rate for $\vert z_k^\top x_l \vert$ than for $\vert x_l^\top w_k \vert$ since we assume existence of the fourth moments.  Then,
\begin{align}\label{eq:rates}
\begin{split}
\dfrac{1}{n}\hat{z}_k^\top \hat{w}_k & = \dfrac{1}{n}z_k^\top P^{\perp}_{-k} w_k =\dfrac{1}{n} \left( z_k^\top w_k - z_k^\top P_{-k} w_k\right) = \dfrac{1}{n}  z_k^\top w_k + { \scriptstyle \mathcal{O}}_p\left(\dfrac{1}{\surd{n}}\right) \\
& = \EE\left(Z_k W_k\right) + { \scriptstyle \mathcal{O}}_p\left(1\right), \\
\dfrac{1}{\surd{n}}\hat{z}_k^\top \hat{w}_k & = \dfrac{1}{\surd{n}}  z_k^\top w_k + { \scriptstyle \mathcal{O}}_p\left(1\right) \overset{\mathbb{D}}{\to} \mathcal{N} \left\{\EE\left(Z_k W_k\right), \text{var}\left(Z_k W_k\right)\right\} \quad \text{and} \\
\dfrac{1}{n}\hat{z}_k^\top \hat{z}_k & = \dfrac{1}{n}z_k^\top P^{\perp}_{-k} z_k =\dfrac{1}{n} \left( z_k^\top z_k - z_k^\top P_{-k} z_k\right) = \dfrac{1}{n}  z_k^\top z_k +  \mathcal{O}_p\left(\dfrac{1}{n}\right) \\
& = \EE\left(Z_k^2\right) + { \scriptstyle \mathcal{O}}_p\left(1\right).
\end{split}
\end{align}
The second line is restricted to covariates for which this variance exists, which includes all non-ancestors as $Z_k \perp W_k$. Using Slutsky's theorem, we have
\begin{align}\label{eq:bhat-conv}
\begin{split}
\hat{\beta}_k^{f,j} & = \dfrac{\EE\left(Z_k W_k\right)}{\EE\left(Z_k^2\right)} + {\scriptstyle\mathcal{O}}_p\left(1\right) = \beta_k^{f,j} + {\scriptstyle \mathcal{O}}_p\left(1\right) \ \forall k \quad \text{and}\\
 \surd{n}\hat{\beta}_k^{f,j} & \overset{\mathbb{D}}{\to} \mathcal{N}\left\{0, \dfrac{\EE\left(W_k^2\right)}{\EE\left(Z_k^2\right)}\right\}\forall k \not \in \left\{\text{AN}\left(j\right)\cup j\right\},
\end{split}
\end{align}
which proves the first part of the theorem.

It remains to consider the variance estimate. Similar to above, we have
\begin{equation*}
n \left(x^\top x\right)^{-1}_{kk} \overset{\mathbb{P}}{\to}\left(\Sigma^X\right)^{-1}_{kk} \equiv \dfrac{1}{\EE\left(Z_k^2\right)}= \mathcal{O}\left(1\right).
\end{equation*}
Further,
\begin{equation*}
\hat{\sigma}^2 = \dfrac{\left\Vert f\left(x_j\right) - x\hat{\beta}^{f,j}\right\Vert_ 2^2}{n-p} \coloneqq \dfrac{1}{n-p}\hat{\eps}^\top\hat{\eps} =\dfrac{1}{n-p}\eps^\top P^{\perp}_{x}\eps=\dfrac{1}{n-p}\left(\eps^\top \eps -\eps^\top P_{x}\eps\right).
\end{equation*}
Similar to before
\begin{align*}
\dfrac{1}{n-p}\vert \eps^\top P_{x}\eps \vert & = \dfrac{1}{n-p} {\scriptstyle \mathcal{O}}_p\left(n\right)\mathcal{O}_p\left(\dfrac{1}{n}\right){ \scriptstyle \mathcal{O}}_p\left(n\right)= { \scriptstyle \mathcal{O}}_p\left(1\right) \quad \text{such that}\\
\hat{\sigma}^2 & = \dfrac{1}{n-p} \eps^\top \eps + { \scriptstyle \mathcal{O}}_p\left(1\right) = \EE\left(\mathcal{E}^2\right) + { \scriptstyle \mathcal{O}}_p\left(1\right) = \mathcal{O}_p\left(1\right).
\end{align*}
Combined, we find
\begin{equation*}
n\widehat{\text{var}}\left(\hat{\beta}^{f,j}_k\right) = \mathcal{O}_p\left(1\right) \leftrightarrow \widehat{\text{var}}\left(\hat{\beta}^{f,j}_k\right) = \mathcal{O}_p\left(\dfrac{1}{n}\right),
\end{equation*}
proving the second part of the theorem.

For non-ancestors, $\beta_k^{f,j} = 0$ such that $W_k = \mathcal{E}$. Then,
\begin{equation*}
n\widehat{\text{var}}\left(\hat{\beta}^{f,j}_k\right) = \dfrac{\EE\left(\mathcal{E}^2\right)}{\EE\left(Z_k^2\right)} + { \scriptstyle \mathcal{O}}_p\left(1\right)= \dfrac{\EE\left(W_k^2\right)}{\EE\left(Z_k^2\right)} + { \scriptstyle \mathcal{O}}_p\left(1\right)
\end{equation*}
such that the last statement of Theorem \ref{theo:norm} follows again from Slutsky's theorem and \eqref{eq:bhat-conv}.

\subsection{Proof of Theoren \ref{theo:adv}}

We generally have the following identity
\begin{equation*}
E \left\{X_l f\left(X_j\right)\right\} = E \left\{E\left(X_l \mid X_j\right) f\left(X_j\right)\right\}.
\end{equation*}

Consider first the simpler case where the $j$-restricted Markov boundary is the full Markov boundary, i.e., all children of $k$ are ancestors of $j$ or $j$ itself.
Let $\Omega \coloneqq E \left(X X^\top \right)^{-1}$ and $\Omega_{kk} \coloneqq d_k$. Then, we have the off-diagonal elements
\begin{equation*}
\Omega_{kl} = \begin{cases}
-d_k \gamma_l^{j,k} & \text{if } l \in \text{MA}^{\rightarrow j}\left(k\right)\\
0 & \text{otherwise}
\end{cases},
\end{equation*}
which is a standard fact from least squares regression. Thus,
\begin{align*}
\beta_k^{f,j} & = \sum_{l=1}^p \Omega_{kl} E\left\{X_l f\left(X_j\right)\right\} = d_k E\left\{X_k f\left(X_j\right)\right\} - d_k \sum_{l \in \text{MA}^{\rightarrow j}\left(k\right)}  \gamma_l^{j,k} E\left\{X_l f\left(X_j\right)\right\} \\
&  = d_k E\left\{E \left(X_k \mid X_j\right) f\left(X_j\right)\right\} - d_k \sum_{l \in \text{MA}^{\rightarrow j}\left(k\right)}  \gamma_l^{j,k} E\left\{E\left(X_l \mid X_j\right) f\left(X_j\right)\right\} \\
& = d_k E \left\{E \left(X_k - \sum_{l \in \text{MA}^{\rightarrow j}\left(k\right)}  \gamma_l^{j,k} X_l \mid X_j \right) f \left(X_j\right) \right\}
%&  = d_k E\left\{E \left(X_{\text{MA}^{\rightarrow j}\left(k\right)}^\top \gamma^{j,k} \mid X_j\right) f\left(X_j\right)\right\} d_k - \sum_{l \in \text{MA}^{\rightarrow j}\left(k\right)}  \gamma_l^{j,k} E\left\{E\left(X_l \mid X_j\right) f\left(X_j\right)\right\} \\
%& =d_k \sum_{l \in \text{MA}^{\rightarrow j}\left(k\right)}  \gamma_l^{j,k} E\left\{E\left(X_l \mid X_j\right) f\left(X_j\right)\right\} - d_k\sum_{l \in \text{MA}^{\rightarrow j}\left(k\right)}  \gamma_l^{j,k} E\left\{E\left(X_l \mid X_j\right) f\left(X_j\right)\right\} = 0.
\end{align*}
This quantity is $0$ for all possible $f\left(\cdot \right)$ iff the conditional expectation is the constant $0$-function. The if-statement is trivial. For the only if, note that one could choose
\begin{equation*}
f \left(X_j\right)  = E \left(X_k - \sum_{l \in \text{MA}^{\rightarrow j}\left(k\right)}  \gamma_l^{j,k} X_l \mid X_j \right)
\end{equation*}
leading to a nonzero expectation unless $f\left(X_j \right) \equiv 0$. Using
\begin{equation*}
E\left(X_k \mid X_j \right) = E\left\{E\left(X_k \mid X_j, X_{\text{MA}^{\rightarrow j}\left(k\right)}\right) \mid X_j \right\}= E\left\{E\left(X_k \mid X_{\text{MA}^{\rightarrow j}\left(k\right)}\right) \mid X_j \right\},
\end{equation*}
the last part of the theorem follows directly.

For the general case, note that for $l$ in the difference between the Markov boundary and the $j$-restricted Markov boundary, $\beta_l^{f,j} = 0$ $\forall f\left(\cdot\right)$ follows from Theorem \ref{theo:beta-fols}. Thus, the least squares parameter for $k$ and $r \in \text{MA}^{\rightarrow j}\left(k\right)$ is the same as if these variables did not exist. Therefore, the result for the general case follows directly from the simpler case discussed above.

\subsection{Proof of Proposition \ref{prop:gauss-psi}}
Consider the least squares solution when only the variables from the restricted Markov boundary are the predictors. 
From Lemma \ref{lemm:zrep}, we get that the residuum, say $\tilde{Z}_k$ is a linear combination of $\Psi_k$ and $\Psi_l$ for $l \in \text{CH}^{\rightarrow j }\left(k\right)$. For every $r \in \text{MA}^{\rightarrow j }\left(k\right)$,
\begin{equation*}
X_r = \sum_{t \in \left\{\text{AN}\left(r\right) \cup r \right\}} \omega_{rt}\Psi_t,
\end{equation*} 
and, dependence with $\tilde{Z}_k$ could only be induced by
\begin{equation*}
\tilde{X}_r = \sum_{t \in \left\{\text{AN}\left(r\right) \cup r \right\} \cap \left\{\text{CH}^{\rightarrow j }\left(k\right) \cup k \right\}} \omega_{rt}\Psi_t.
\end{equation*} 
By the least squares property, $\tilde{X}_r$ and $\tilde{Z}_k$ are uncorrelated. By the Gaussianity of $\Psi_k$ and $\Psi_l$, this implies independence. Thus, $\tilde{Z}_k$ is independent from all $X_r$ such that the linear least squares fit is also the conditional expectation. Thus, the sufficient condition from Theorem \ref{theo:adv} for $\beta_k^{f,j} = 0$ holds.

Due to Gaussianity and Lemma \ref{lemm:zrep}, $Z_k$ is independent from $\text{CH}^{\rightarrow j }\left(k\right)$, their descendants, and all its non-descendants. Therefore, it is also independent from
\begin{equation*}
\mathcal{E} = W_k = f\left(X_j\right) - X_{\left\{\text{AN}\left(j\right) \cup j \right\} \setminus k}\beta_{\left\{\text{AN}\left(j\right) \cup j \right\} \setminus k}^{f,j}
\end{equation*}
such that the variance as in \eqref{eq:rates} is consistently estimated.

\subsection{Proof of Proposition \ref{prop:distr}}
Decompose the conditional expectation as
\begin{align*}
E\left(X_k \mid X_{\text{MA}^{\rightarrow j }\left(k\right)}\right) & = E\left(\Psi_k + \sum_{r \in \text{PA}\left(k\right)} \theta_{k,r} X_r \mid X_{\text{MA}^{\rightarrow j }\left(k\right)}\right)=E\left(\Psi_k \mid X_{\text{MA}^{\rightarrow j }\left(k\right)}\right) + \sum_{r \in \text{PA}\left(k\right)} \theta_{k,r} X_r
%& = E\left(\Psi_k \mid X_{l \cup \text{PA}\left(l\right) \setminus k}\right) + \sum_{r \in \text{PA}\left(k\right)} \theta_{k,r} X_r.
\end{align*}
Recall the definition $\text{CH}^{\rightarrow j}\left(k\right)=\{l\}$. Then,
\begin{align*}
X_{l} &  = E\left(X_{l} \mid X_{l}\right)= E\left(X_{l} \mid X_{\text{MA}^{\rightarrow j }\left(k\right)}\right) = E\left(\Psi_l + \theta_{l,k} X_k + \sum_{t \in \text{PA}\left(l\right) \setminus k} \theta_{l,t} X_t  \mid X_{\text{MA}^{\rightarrow j }\left(k\right)}\right) \\
& =  E\left(\Psi_l + \theta_{l,k} \Psi_k + \theta_{l,k}\sum_{r \in \text{PA}\left(k\right)} \theta_{k,r} X_r + \sum_{t \in \text{PA}\left(l\right) \setminus k} \theta_{l,t} X_t  \mid X_{\text{MA}^{\rightarrow j }\left(k\right)}\right)\\
& = E\left(\Psi_l + \theta_{l,k} \Psi_k   \mid X_{\text{MA}^{\rightarrow j }\left(k\right)}\right) + \theta_{l,k}\sum_{r \in \text{PA}\left(k\right)} \theta_{k,r} X_r + \sum_{t \in \text{PA}\left(l\right) \setminus k} \theta_{l,t} X_t
\end{align*}
such that
\begin{align*}
X_l - \theta_{l,k}\sum_{r \in \text{PA}\left(k\right)} \theta_{k,r} X_r - \sum_{t \in \text{PA}\left(l\right) \setminus k} \theta_{l,t} X_t & = E\left(\Psi_l + \theta_{l,k} \Psi_k   \mid X_{\text{MA}^{\rightarrow j }\left(k\right)}\right)  = 2 \theta_{l,k} E\left(\Psi_k \mid X_{\text{MA}^{\rightarrow j }\left(k\right)}\right).
\end{align*}
The last equality follows since $\Psi_l \overset{\mathbb{D}}{=} \theta_{l,k} \Psi_k$ and both random variables depend on the conditioning set on the same way. Therefore, all the terms in 
$E\left(X_k \mid X_{\text{MA}^{\rightarrow j }\left(k\right)}\right)$ are linear combination such that the sufficient condition from Theorem \ref{theo:adv} holds.

However, $Z_k \not \perp X_l$ in general such that $\text{var}\left(Z_k W_k\right) = \EE\left(Z_k^2\right)\EE\left(W_k^2\right)$ is not generally true. Then, the limiting distribution of the estimator is not the same as for non-ancestors; see also \eqref{eq:rates}.

\subsection{Proof of Corollary \ref{corr:algo}}
Let $j,k$ be such that $k \in \widehat{\text{AN}}\left(j\right)$. This means that there is at least one set\\
$M=\left\{m_0 = k, m_1, \ldots, m_{t-1}, m_t=j\right\}$ such that $P_{m_{s-1}}^{m_s} < \alpha \ \forall s \in \left\{1, \ldots, t\right\}$, where $t\geq 1$. If $k \not\in \text{AN}\left(j\right)$ at least one of these must correspond to a false causal discovery, i.e., there is an $s$ such that $P_{m_s-1}^{m_s} < \alpha$ but $m_{s-1} \not \in \text{AN}\left(m_s\right)$. We conclude
\begin{equation*}
\left\{\exists j,k\neq j: \ k \not\in \text{AN}\left(j\right) \text{ and } k \in \widehat{\text{AN}}\left(j\right)\right\} \rightarrow \left\{\exists j,k \neq j: \ k \not\in \text{AN}\left(j\right) \text{ and } P_k^j<\alpha\right\}.
\end{equation*}
Let $r = \sum_{j,k \neq j} 1_{\left\{H_{0,k \rightarrow j} \text{ is true}\right\}}$ denote the number of true null hypotheses. By the construction of Bonferroni-Holm
\begin{equation*}
\left\{\exists j,k \neq j: \ k \not\in \text{AN}\left(j\right) \text{ and } P_k^j<\alpha\right\} \rightarrow \left\{\exists j,k \neq j: \ k \not\in \text{AN}\left(j\right) \text{ and } p_k^j<\alpha/r\right\}.
\end{equation*}
Let $z_k^j = \hat{\beta}^{f,j}_k /\surd{\widehat{\text{var}}\left(\hat{\beta}^{f,j}_k\right)}$ as used in Theorem \ref{theo:norm}. We find
\begin{align*}
& \underset{n \rightarrow \infty}{\text{lim}}\PP\left\{\exists j,k\neq j: \ k \not\in \text{AN}\left(j\right\} \text{ and } k \in \widehat{\text{AN}}\left(j\right) \right)  \leq \underset{n \rightarrow \infty}{\text{lim}}\PP\left\{\exists j,k \neq j: \ k \not\in \text{AN}\left(j\right) \text{ and } P_k^j<\alpha \right\} \\ 
\leq & \underset{n \rightarrow \infty}{\text{lim}}\PP\left\{\exists j,k \neq j: \ k \not\in \text{AN}\left(j\right) \text{ and } p_k^j<\alpha/r \right\} = \underset{n \rightarrow \infty}{\text{lim}}\PP\left(\underset{j,k \neq j: \ k \not\in \text{AN}\left(j\right)}{\text{min}}p_k^j<\alpha/r \right) \\
\leq & \underset{n \rightarrow \infty}{\text{lim}} \sum_{j,k \neq j: \ k \not\in \text{AN}\left(j\right)} \PP\left(p_k^j<\alpha/r \right) = \sum_{j,k \neq j: \ k \not\in \text{AN}\left(j\right)} \underset{n \rightarrow \infty}{\text{lim}} \PP\left(p_k^j < \alpha/r \right) \\
= &\sum_{j,k \neq j: \ k \not\in \text{AN}\left(j\right)} \underset{n \rightarrow \infty}{\text{lim}} \PP\left\{\Psi\left(\vert z_j^k\vert\right)  > 1 - \alpha/2r \right\} =\sum_{j,k \neq j: \ k \not\in \text{AN}\left(j\right)} \underset{n \rightarrow \infty}{\text{lim}} \PP\left\{\vert z_j^k\vert  >  \Psi^{-1}\left(1 - \alpha/2r\right) \right\} \\
= & \sum_{j,k \neq j: \ k \not\in \text{AN}\left(j\right)} 1- \underset{n \rightarrow \infty}{\text{lim}} \PP\left\{\vert z_j^k\vert  \leq  \Psi^{-1}\left(1 - \alpha/2r\right) \right\} = \sum_{j,k \neq j: \ k \not\in \text{AN}\left(j\right)} \alpha/r = \alpha,
\end{align*}
which proves the first part of the corollary. The second to last equality uses Theorem \ref{theo:norm} and the continuous mapping theorem.

As the model \eqref{eq:SEM-model} excludes the possibility of directed cycles, any output of $\textsc{BuildRecursive}$ that contains cycles must include at least one false causal detection. If $\hat{\alpha} < \alpha$, it corresponds to the maximal p-value such that including the corresponding ancestor relationship creates cycles. Therefore, it must hold $\underset{j,k \neq j: \ k \not\in \text{AN}\left(j\right)}{\text{min}}P_k^j \leq \hat{\alpha}$ such that
\begin{equation*}
\underset{n \rightarrow \infty}{\text{lim}} \PP\left(\hat{\alpha} \leq \alpha' \right) \leq \underset{n \rightarrow \infty}{\text{lim}}\PP\left(\underset{j,k \neq j: \ k \not\in \text{AN}\left(j\right)}{\text{min}}P_k^j \leq \alpha' \right) \leq \alpha'
\end{equation*}
using similar arguments as above.

\newpage
\section{Algorithm}\label{app:algo}
\begin{algorithm}[h!]
\caption{Nodewise and recursive ancestor detection}\label{alg:graph}
\hspace*{\algorithmicindent} \textbf{Input} data $x \in \mathbb{R}^{n \times p}$, significance level $\alpha \in \left(0,1\right)$ and nonlinear function $f\left(\cdot\right)$\\
\hspace*{\algorithmicindent} \textbf{Output} Estimated set of ancestors $\widehat{\text{AN}}\left(j\right) \ \forall j \in \left\{1,\ldots,p\right\}$, adjusted significance level $\hat{\alpha}$
\begin{algorithmic}[1]
\For{$j=1$ to $p$}
\State Calculate $p_k^j \ \forall k \neq j$ using \eqref{eq:hat-def} and \eqref{eq:z-test} \textit{\# Calculate the p-values of ancestor regression}
\EndFor
\State Apply a multiplicity correction to the list of $p_k^j \ \forall j, k \neq j$ denote the corrected p-values by $P_k^j$
\State \textit{\# Store p-values in a matrix, descendants as rows, ancestors as columns}
\State Define $P\in \mathbb{R}^{p \times p}$ such that $P_{j,k}=P^j_k$ and $P_{j,j}=1$
\State $\left(A,\ \hat{\alpha}\right) \gets$ \Call{FindStructure}{$P$, $\alpha$}
\For{$j=1$ to $p$}
\State \textit{\# Transform binary ancestor matrix to a list of ancestors for each node}
\State $\widehat{\text{AN}}\left(j\right) \gets \left\{k: A_{j,k} = \texttt{TRUE}\right\}$
\EndFor
\Procedure{FindStructure}{$P\in \mathbb{R}^{d \times d}$, $\alpha$}
\State \textit{\# Define ancestors based on significant p-values}
\State Define $A\in \mathbb{R}^{d \times d}$ such that $A_{j,k} = \texttt{TRUE}$ if $P_{j,k} < \alpha$ and else $A_{j,k} = \texttt{FALSE}$
\State \textit{\# Recursively complete the ancestral sets such that ancestors' ancestors are ancestors}
\State $A \gets$\Call{BuildRecursive}{$A$}
\State $I \gets \left\{j \in \left\{1,\ldots,d\right\}: \ A_{j,j} = \texttt{TRUE}\right\}$ \textit{\# Find nodes that lead to cycles}
\If{$I = \emptyset$} 
\State \Return $\left(A,\ \alpha\right)$ \textit{\# If no cycles remain, output the result of the current significance level}
\Else
\State $\hat{\alpha} \gets \underset{j \in I, k \neq j \in I: \ P_{j,k}<\alpha}{\text{max}}P_{j,k}$ \textit{\# Otherwise, reduce $\alpha$ to remove at least one edge}
\State $\left(A_{I,I},\ \hat{\alpha}\right)\gets$ \Call{FindStrucure}{$P_{I,I}$, $\hat{\alpha}$} \textit{\# Find structure for variables in cycles}
\State $A \gets$\Call{BuildRecursive}{$A$} \textit{\# Once no more cycles occur, complete the ancestral sets}
\State \Return $\left(A,\ \hat{\alpha}\right)$
\EndIf
\EndProcedure
\Procedure{BuildRecursive}{$A\in \mathbb{R}^{d \times d}$}
\For{$j=1$ to $d$}
\State $\widehat{\text{AN}}\left(j\right) \gets \left\{k: A_{j,k} = \texttt{TRUE}\right\}$ \textit{\# Initiate ancestors based on p-values}
\EndFor
\For{$j=1$ to $d$}
\State $S \gets \emptyset$ \textit{\# Set of ancestors that have been checked, initiated as empty}
\While{$\widehat{\text{AN}}\left(j\right) \setminus S \neq \emptyset$}
\For{$k \in \widehat{\text{AN}}\left(j\right) \setminus S$}
\State \textit{\# Add ancestors' ancestors until all are checked}
\State $\widehat{\text{AN}}\left(j\right) \gets \widehat{\text{AN}}\left(j\right) \cup \widehat{\text{AN}}\left(k\right)$ and $S \gets S \cup K$
\EndFor
\EndWhile
\State $A_{j,\widehat{\text{AN}}\left(j\right)} \gets \texttt{TRUE}$ \textit{\# Store to matrix format}
\EndFor
\State \Return $A$
\EndProcedure
\end{algorithmic}
\end{algorithm}

\section{Details on the simulation setup}\label{app:sim}
We use the following distributions for the $\Psi_j$: two $t_7$ distributions, a centered Laplace distribution with scale $1$, a centered uniform distribution, and a standard normal distribution. Depending on the scenario, the last error distribution is either uniform or Gaussian. The results in Section \ref{sim:anc} are from the former case. All distributions are normalized to having unit variance. For each simulation run, we randomly permute the distributions to assign them to $\Psi_1$ to $\Psi_6$.

We create an edge between the two variables with (potentially) Gaussian error term. The remaining $14$ edges $X_k \rightarrow X_j$ with $k<j$ are present with probability $5/14$ each such that an average of $6$ parental connections exists.

We assign preliminary edge weights uniformly in $\left[0.5, 1\right]$. These are further scaled such that for every $X_j$ which is not a source node, the standard deviation of
\begin{equation*}
    \sum_{k \in \text{PA}\left(j\right)}\theta_{j,k} X_k
\end{equation*}
is uniformly chosen from $\left[\surd 0.5, \surd 2\right]$. Thus, the signal-to-noise ratio is between $1/2$ and $2$.

To initialize the graph and the weights, we use the function \texttt{randomDAG} from the \texttt{R}-package \texttt{pcalg} \citep{kalisch2012causal} before applying our changes to enforce the constraints.
\section{Additional simulation results}\label{app:sim-mix}
\begin{figure}[t!]
%\spacingset{1}
 \centering
 \includegraphics[width=0.8\textwidth]{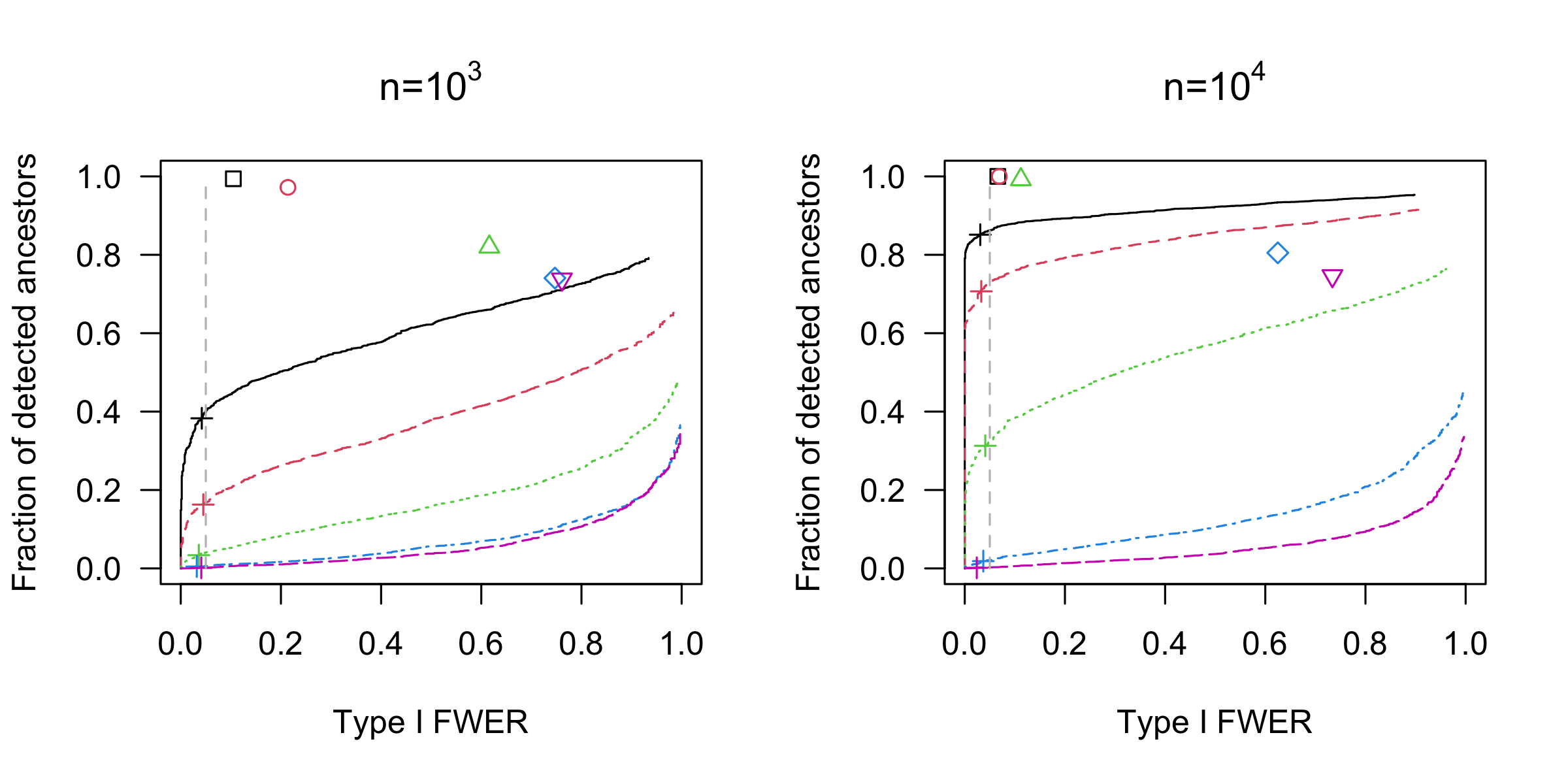}
 \caption[Simulation in a linear SEM]
 {Nodewise ancestor detection in a linear structural equation model with $6$  variables. The results are based on $1000$ simulation runs. Depicted is the family-wise error rate of false causal detection versus the fraction of detected ancestors. The curves use the level of the test $\alpha$ as implicit curve parameter. The pluses correspond to nominal $\alpha = 5\%$. The vertical line is at actual $5\%$. The other symbols correspond to the performance of the LiNGAM algorithm. We consider the different values of $\gamma$: $0$ (solid / square, black), $0.25$ (dashed / circle, red), $0.5$ (dotted / triangle pointing upward, green), $0.75$ (dot-dashed / diamond, blue), and $1$ (long-dashed / triangle pointing downward, pink). The sample size is $10^3$ on the left and $10^4$ on the right.}
 \label{fig:graph-perf-mix}
\end{figure}

To analyse the effect of close to Gaussian error distributions we consider a further variation of the first the scenario in \ref{app:sim}. Call the normalized error terms from before $\Psi_j'$. These are mixed with a standard Gaussian component $\Psi_j''$ such that
\begin{equation*}
\Psi_j = \sqrt{1 - \gamma} \Psi_j' + \sqrt{\gamma} \Psi_j'' \ \forall j.
\end{equation*}
Thus, the Gaussian term causes a fraction $\gamma$ of the variance. We vary $\gamma$ from $0$, which is the setup from before, to $1$ in steps of $0.25$. We consider the same performance metrics as in Figure \ref{fig:graph-perf-rand}.  For the sake of overview, we restrict ourselves to $n=10^3$ and $n=10^4$ in Figure \ref{fig:graph-perf-mix}.

For both LiNGAM and ancestor regression, increasing the amount of Gaussianity leads to a performance drop. Thus, not only fully Gaussian error terms harm these methods. For $\gamma = 0.75$, $10^4$ samples are not sufficient to keep the type I error of LiNGAM low. For ancestor regression, nearly Gaussian error distribution leads to a substantial drop in power. However, the type I error remains under control supporting Corollary \ref{corr:algo}. While power considerations are clearly in favor of LiNGAM, especially in easy scenarios, our method leads to fewer but more trustworthy findings in close to unidentifiable scenarios within the class of linear structural equation models.

\vspace*{20in}
\end{document}